# The angular correlation hierarchy in the quasilinear regime

F. Bernardeau[1,2]

[1] Service de Physique Théorique*, CE de Saclay, F-91191 Gif-sur-Yvette Cedex, France (present address)
[2] CITA, University of Toronto, 60 St. George St., Toronto, ON, Canada M5S 1A1

**Abstract.** For Gaussian initial conditions the perturbation theory predicts a very specific hierarchy for the projected matter $p$-point correlation functions. In the small angle approximation and assuming a power-law spectrum I derive the exact expressions of the coefficients $s_p$ relating the averaged $p$-order angular correlation function, $\overline{\omega}_p$ to the second one, $\overline{\omega}_p = s_p\,\overline{\omega}_2^{p-1}$. These results are valid for any selection function, but for a top-hat angular filter only. These coefficients are found to be significantly higher than their 3D counterparts, $S_p = \overline{\xi}_p/\overline{\xi}_2^{p-1}$.

For the coefficient $s_3$ I discussed the accuracy of the small angle approximation by computing, for particular examples, its angular dependence with Monte-Carlo numerical integrations. It is found that the accuracy of the small angle approximation for $\theta \approx 1^\circ$ slightly depends on the selection function. Using the selection function expected for galaxy catalogues the approximation is found to be reasonably good.

The measurements of the $s_p$ parameters made in the APM angular survey are found to give systematic lower values than the theoretical predictions. How significant this discrepancy is and what the implications would be for galaxy formation models is discussed in the last section.

## 1. Introduction

The study of the statistical properties of the galaxy distribution had been done initially in angular surveys such as the Lick catalogue (Shane & Wirtanen 1967, Seldner et al. 1977). The more recent apparition of redshift surveys led people to study in more details of the galaxy 3D distribution, and most of the results obtained in 2D catalogues have been confirmed. In particular the index of the 2-point correlation function initially determined by Limber (1954), Neyman, Scott & Shane (1956) and Groth & Peebles (1977) has been reobtained by the analysis of the redshift surveys.

However, due to the huge number of objects contained in angular surveys, they still allow the most precise statistical studies. In particular one can more reliably measure high-order correlation functions in 2D catalogues than in 3D. The APM Galaxy Survey with more than 1,300,000 objects is at present the largest catalogue available and Gaztañaga (1994) was able to measure the averaged galaxy angular $p$-point correlation functions up to $p = 9$. Similar results are beyond the present status of the observations for redshift surveys (Bouchet et al. 1993). Moreover in 2D catalogues there are no redshifts effects that can mix the density field and the velocity field properties, which makes comparisons with theoretical predictions more difficult. The main difficulty brought by the analysis of the angular surveys is that a given angular scale corresponds to various physical scales. Limber (1954) pioneered the expected relationship between the angular 2-point correlation function and the real space correlation function. This relation, and its approximation for small angles (eg., Peebles 1980, §50), proved to be efficient and accurate.

In the strongly nonlinear regime, our theoretical knowledge of the quantitative behaviour of the density field is still weak. $N$-body simulations have been widely used to improve our understanding of this complicated part of the dynamics. The size of the catalogues involved in 2D measurements, however, still challenges the largest simulations. On the other hand, rapid progress have been made recently in the mildly nonlinear regime where perturbation theory can be applied. This led to a better understanding of the early departure from the linear behaviour. In particular it has been shown that in such a regime, the mean value, $\overline{\xi}_p$, of the $p$-point correlation function $\xi_p(\boldsymbol{x}_1,\ldots,\boldsymbol{x}_p)$ in a spherical cell is proportional to $\overline{\xi}_2^{p-1}$,

$$\overline{\xi}_p = S_p\,\overline{\xi}_2^{p-1}, \qquad (1)$$

---

* Laboratoire de la Direction des Sciences de la Matière du Commissariat à l'Energie Atomique

where $S_p$ are coefficients depending on the shape of the power spectrum only (Fry 1984, Goroff et al. 1986, Juszkiewicz, Bouchet & Colombi 1993, Bernardeau 1994a, b). This hierarchy is specific of Gaussian initial conditions. For non-Gaussian initial conditions similar hierarchy may or may not be obtained. It seems to depend explicitly on the specificities of the model. Early studies (Jaffe 1994) tend to demonstrate that the hierarchy (1) is a robust result and holds even for mildly non-Gaussian initial conditions. Numerical studies show that the analytical results obtained in the quasilinear regime are very accurate to describe the correlation hierarchy up to $\sigma = 1$ (Juszkiewicz et al. 1994, Bernardeau 1994, Lokas et al. 1994, Baugh, Gaztañaga & Efstathiou 1994). The 3D statistics of the one-point density PDF in the quasilinear regime has thus been investigated in details, but for the reasons mentionned before comparisons with the observations remain questionable. It may be worth to note that numerical simulations show that the nonlinear corrections tend to increase the value of the coefficients $S_p$ (Bouchet & Hernquist 1992, Colombi et al. 1994, Lucchin et al. 1994).

However, so far, substantial progress has been made only for the statistical properties of the density field when it is averaged in a spherical cell. The properties of a smoothed angular density field corresponds rather to a filtering in a cone of a given solid angle. It implies that, for 2D catalogues, the connected part of the moments, the cumulants $c_p$ (for a definition, see Balian & Schaeffer 1989), are expected to exhibit a similar specific behaviour (Schaeffer 1987),

$$c_p = s_p \, c_2^{p-1}, \qquad (2)$$

but with different coefficients $s_p$. These coefficient $s_p$ are expected to depend on the power spectrum as well as on the shape of the selection function, $F(r)$. Note that the property (2) can be expressed in term of the $p$-point angular correlation function, $w_p(\gamma_1, \ldots, \gamma_p)$. The relation (2) implies that the average of the $p$-point angular correlation functions, $\overline{w}_p$, within a cell of angular radius $\theta_0$ follow the hierarchy,

$$\overline{w}_p = s_p \, \overline{w}_2^{p-1}. \qquad (3)$$

The hierarchy (3) is indeed found to describe accurately the statistics of both the galaxies and the galaxy clusters. The parameters $s_p$ are measurable up to $p = 6$ for the clusters and up to $p = 9$ at small scales for the galaxies (Cappi & Maurogordato 1994, Gaztañaga 1994). This result has been used as an argument in favor of the gravitational instability scenario and for the absence of biases in the galaxy distribution. Unfortunatly the parameters $S_p$, corresponding to another smoothing scheme, cannot be simply derived from the observations in 2D catalogues. The approximation used by Tóth, Hollósi and Szalay, (1989), and Gaztañaga, (1994), has been improperly used since it is valid only for exact tree models for the $p$-point correlation functions. Therefore, it should not be used for perturbation theory calculations. It gives however an estimation of these coefficients that is attractively close to the measured values (Gaztañaga 1994, Frieman & Gaztañaga 1994). It is then worth to derive more exactly the quantitative behaviour of the high order cumulants from the perturbation theory.

Throughout this paper, I make the hypothesis that the galaxies are good tracers of the matter field. In §2 I present the expected values of the coefficients $s_p$ as given by the perturbation theory in the small angle approximation. §3 is devoted to a discussion on the accuracy of the small angle approximation on the value of $s_3$ and the last section is devoted to a comparison of these results with observations and to comments.

## 2. The projection effects in perturbation theory: the correlation hierarchy in the small angle approximation

### 2.1. Physical assumptions

The basic physical assumption is that the local number density of observable objects, $n(\boldsymbol{x})$, is assumed to be proportional to the local matter density $\rho(\boldsymbol{x})$ multiply by a selection function which is function of distance, $F(r)$,

$$n(\boldsymbol{x}) = n_0 \, F[|\boldsymbol{x}|] \, \rho(\boldsymbol{x})/\rho_0, \qquad (4)$$

where $\rho_0$ is the mean density of the Universe. The quantity $n_0 F(r)$ is then the mean density of *observable* objects at the distance $r$. The selection function obviously depends on the sample that is considered. In the text I discuss the results for different shape of selection functions.

The projected number density of objects, $\sigma_0$, per solid angle is

$$\sigma_0 \, \mathrm{d}\Omega = \mathrm{d}\Omega \, n_0 \int_0^\infty r^2 \mathrm{d}r F(r), \qquad (5a)$$

and the local angular density, at the angular position defined by the unit vector $\boldsymbol{\gamma}$, is

$$\sigma(\boldsymbol{\gamma}) = n_0 \int_0^\infty r^2 \mathrm{d}r F(r) \rho(r\boldsymbol{\gamma})/\rho_0. \qquad (5b)$$

The correlation properties of the angular density then depends on the hypothesis made for the matter density field. The complete series of the angular correlation functions, $\omega_p(\boldsymbol{\gamma}_1, \ldots, \boldsymbol{\gamma}_p)$, are difficult to measure. What can be more confidently measured are the spatial averages of these functions in a cell of angular size $\theta_0$,

$$\overline{\omega}_p = \int \mathrm{d}^2 \boldsymbol{\gamma}_1 \, W(\boldsymbol{\gamma}_1) \ldots \mathrm{d}^2 \boldsymbol{\gamma}_p \, W(\boldsymbol{\gamma}_p) \omega_p(\boldsymbol{\gamma}_1, \ldots, \boldsymbol{\gamma}_p) \qquad (6)$$

where $W(\boldsymbol{\gamma})$ is unity if the angle between $\boldsymbol{\gamma}$ and a given direction $\boldsymbol{\gamma}_0$ is less than $\theta_0$ and zero otherwise (the values of $\overline{\omega}_p$, according to the isotropic principle, should not depend on the direction $\gamma_0$). These averages are simply given by the cumulants of the number density probability distribution function when the angular field is filtered by a top-hat filter of radius $\theta_0$.

The last important assumptions are that, at sufficiently large scale, the matter correlation properties can be described with the perturbation theory; and that the initial conditions were Gaussian. In most of the following calculations I will also assume that these scales are reached in the present 2D catalogues even for small filtering angles, so that for these calculations I will assume that

$$\theta_0 \ll 1. \tag{7}$$

The validity of this assumption is discussed in §3.

The assumption that perturbation theory can described the matter correlation properties implies that the three-point correlation function, for instance, has to be calculated using both the $1^{\rm st}$ and the $2^{\rm nd}$ order for the density field in perturbation theory. The properties of the linear density field are entirely given by the shape of the power spectrum $P(k)$. It is defined in such a way that the Fourier transforms of the linear density contrast, given by

$$\rho^{(1)}(\boldsymbol{x})/\rho_0 = \int \frac{{\rm d}^3\boldsymbol{k}}{(2\pi)^{3/2}} \delta(\boldsymbol{k}) \exp({\rm i}\boldsymbol{k}\boldsymbol{x}),$$

are Gaussian variables of moments

$$\langle \delta(\boldsymbol{k}_1)\delta(\boldsymbol{k}_2)\rangle = \delta_{\rm dirac}(\boldsymbol{k}_1+\boldsymbol{k}_2)P(k_1). \tag{8}$$

In the following the power spectrum is approximated by a power law,

$$P(k) = A k^n, \tag{9}$$

normalized to the present time. The index $n$ can in principle be determined from the shape of the angular two-point correlation function.

2.2. The variance and the skewness

To start with let me remind the expression of the density field at the $2^{\rm nd}$ order,

$$\frac{\rho^{(2)}(\boldsymbol{x})}{\rho_0} = \int \frac{{\rm d}^3\boldsymbol{k}_1}{(2\pi)^{3/2}}\frac{{\rm d}^3\boldsymbol{k}_2}{(2\pi)^{3/2}} \delta(\boldsymbol{k}_1) \delta(\boldsymbol{k}_2) \exp[{\rm i}(\boldsymbol{k}_1+\boldsymbol{k}_2)\boldsymbol{x}] \\ \times \left[\frac{5}{7} + \frac{\boldsymbol{k}_1.\boldsymbol{k}_2}{k_2^2} + \frac{2}{7}\left(\frac{\boldsymbol{k}_1.\boldsymbol{k}_2}{k_2\,k_1}\right)^2\right]. \tag{10}$$

This expression is exact for an Einstein-de Sitter Universe only. This cosmological hypothesis is assumed in §2.2 and 2.3. The dependence of the results with $\Omega$ is discussed in §2.4.

As the expression of the $2^{\rm nd}$ order density field is simple only in Fourier space we are led to derive the expression of the variance in this space. The first step is to rederive the small angle approximation for the variance (eg. Peebles 1980, §50) with such an approach. To do so I use the fact that, when $\theta_0 \ll 1$,

$$\int {\rm d}^2\boldsymbol{\gamma}\, W(\boldsymbol{\gamma})\, \exp({\rm i}\,x\,\boldsymbol{k}.\boldsymbol{\gamma}) \approx \\ 2\pi\,\theta_0^2\, \exp[{\rm i}\,x\,k\,\cos(\theta_k)]\frac{J_1[k\,x\,\sin(\theta_k)\,\theta_0]}{k\,x\,\sin(\theta_k)\,\theta_0}, \tag{11}$$

where $k$ is the modulus of $\boldsymbol{k}$, $\theta_k$ is the angle between $\boldsymbol{k}$ and $\boldsymbol{\gamma}_0$ and $J_1$ is the Bessel function of the first kind. Using the relation (11) it is possible to derive the expression of the variance, $\overline{\omega}_2$,

$$\overline{\omega}_2 = \frac{1}{\left[\int_0^\infty r^2 {\rm d}r F(r)\right]^2} \int_0^\infty r_1^2\,{\rm d}r_1\,F(r_1)\,\int_0^\infty r_2^2\,{\rm d}r_2\,F(r_2) \\ \times \int_{-\infty}^{+\infty} \frac{{\rm d}k_r}{2\pi} \int \frac{{\rm d}^2\boldsymbol{k}_\perp}{(2\pi)^2} P(k)\, \exp[{\rm i}\,k_r\,(r_1-r_2)] \\ \times W_2(k_\perp\,r_1\,\theta_0)\,W_2(k_\perp\,r_2\,\theta_0) \tag{12}$$

where I wrote

$$\boldsymbol{k} = k_r\boldsymbol{\gamma}_0 + \boldsymbol{k}_\perp \tag{13}$$

and

$$W_2(x) = 2\frac{J_1(x)}{x}. \tag{14}$$

In the limit of small angle $\theta_0$, the cell in which the integration is made is extremely elongated so that the component of $k$ along the line of sight is negligible compared to its orthogonal component, which implies

$$k \approx k_\perp. \tag{15}$$

With such an approximation the integrale (12) simplifies dramatically. Indeed the integration over $r$ is straightforward and implies that $r_1 = r_2$ so that (12) reads,

$$\overline{\omega}_2 = \frac{A\,\theta_0^{-2-n}}{2\pi} \frac{\int_0^\infty r^{2-n}{\rm d}r F^2(r)}{\left[\int_0^\infty r^2 {\rm d}r F(r)\right]^2} \int_0^\infty {\rm d}l\,l^{1+n}\,W_2^2(l). \tag{16}$$

It is possible to show that the expression (16) is consistent with the Limber equation,

$$w(\theta) \propto \int_0^\infty r^4\,F^2(r){\rm d}r \int_{-\infty}^\infty \xi_2\left(\sqrt{x^2+r^2\theta^2}\right){\rm d}x,$$

showing that (15) is indeed equivalent to the usual small angle approximation.

The derivation of the skewness of the angular density field is based on similar approximations. It involves the integration over two different wave vectors, $\boldsymbol{k}_1$ and $\boldsymbol{k}_2$. Their components along $\boldsymbol{\gamma}_0$ are then also negligible compared to their orthogonal components. As a result, in the expression of the $2^{\rm nd}$ order density field the geometrical part $5/7+\boldsymbol{k}_1.\boldsymbol{k}_2/k_2^2+2/7\,(\boldsymbol{k}_1.\boldsymbol{k}_2/k_2\,k_1)^2$ involves only the averaged three-point correlation function then reads,

$$\overline{\omega}_3 = \frac{6A^2\, \theta_0^{-2(n+2)}}{(2\pi)^4} \frac{\int_0^\infty r^{8-2(n+3)} \mathrm{d}r\, F^3(r)}{\left[\int_0^\infty r^2 \mathrm{d}r\, F(r)\right]^3} \int_0^\infty \mathrm{d}^2 l_1\, \mathrm{d}^2 l_2$$
$$\times\, l_1^{1+n}\, W_2(l_1)\, l_2^{1+n}\, W_2(l_2)\, W_2(|l_1 + l_2|) \quad (17)$$
$$\times\, \left[5/7 + l_1.l_2/l_2^2 + 2/7\,(l_1.l_2/l_2\, l_1)^2\right].$$

The calculation of the integrale over the vectors $l_1$ and $l_2$ turns out to lead to simple integrations because of special geometrical properties of the function $W_2$. Indeed we have

$$\int_0^{2\pi} \mathrm{d}\varphi\, W_2(|l_1 + l_2|)\, [1 - \cos^2(\varphi)] = \pi\, W_2(l_1)\, W_2(l_2) \quad (18)$$

where $\varphi$ is the angle between $l_1$ and $l_2$. We can also notice that

$$\frac{2}{l}\frac{\mathrm{d}J_0(l)}{\mathrm{d}l} = -W_2(l) \quad (19)$$

and that

$$\int_0^{2\pi} \mathrm{d}\varphi\, J_0(|l_1 + l_2|) = 2\pi\, J_0(l_1)\, J_0(l_2) \quad (20)$$

which by differentiating with respect to $l_2$ leads to the relationship

$$\int_0^{2\pi} \mathrm{d}\varphi\, W_2(|l_1 + l_2|)\, [1 + \cos(\varphi)l_1/l_2]$$
$$= 2\pi\, J_0(l_1)\, W_2(l_2) \quad (21)$$
$$= 2\pi\, W_2(l_2) \left[W_2(l_1) + \frac{l_1}{2}W_2'(l_1)\right].$$

The properties (18) and (20) can be shown using similar decompositions (in sum of product of Bessel functions) than the ones used for the Fourier transform of the top-hat window function in 3D (Bernardeau 1994a). The demonstrations will not be given.

Using the relationships (18) and (21) one can then calculate the expression (17). The result reads

$$s_3 \equiv \frac{\overline{\omega}_3}{\overline{\omega}_2^2} = R_3\,\left[\frac{36}{7} - \frac{3}{2}(n+2)\right] \quad (22)$$

with

$$R_3 = \frac{\int_0^\infty r^{8-2(n+3)} \mathrm{d}r\, F^3(r)\, \int_0^\infty r^2 \mathrm{d}r\, F(r)}{\left[\int_0^\infty r^{5-(n+3)} \mathrm{d}r\, F^2(r)\right]^2}. \quad (23)$$

The expression of the measured coefficient $s_3$ then reduced to a simple expression where its dependence with the power spectrum index, and the shape of the selection function can be easily investigated.

## 2.3. High-order correlations

The generalization of the results obtained in part 2.2 to high-order correlation function is actually quite straightforward. Let me remind the form of the general expression of the density contrast at the $p^{\mathrm{th}}$ order (Goroff et al. 1986),

$$\frac{\rho^{(p)}}{\rho_0} = \int \frac{\mathrm{d}^3 k_1}{(2\pi)^{3/2}} \ldots \frac{\mathrm{d}^3 k_p}{(2\pi)^{3/2}}\, \delta(k_1)\ldots\delta(k_p)$$
$$\times N_p(k_1,\ldots,k_p) \quad (24)$$

where $N_p$ is a homogeneous function of the wave vectors $k_1,\ldots,k_p$.

The $p$-order correlation function in the quasi-linear regime is given by product of the density field at various order following a kind of tree form (Fry 1984, Bernardeau 1992). When one applies the approximation (15) for each of the wave vectors involved in the tree product the coefficient $s_p$ is expected to take the form (Tóth et al. 1989, Gaztañaga 1994)

$$s_p = R_p\, \Sigma_p(n) \quad (25)$$

where $\Sigma_p$ is a function of $n$ only and

$$R_p = \frac{M_1^{p-2}(3)\, M_p[3p - (p-1)(n+3)]}{M_2^{p-1}(3-n)} \quad (26a)$$

with

$$M_p(a) = \int_0^\infty \mathrm{d}r\, r^{a-1}\, F^p(r). \quad (26b)$$

This is a key property which implies that, in the limit of small angles, the dependence with the shape of the selection function factorizes away and is very simple. Note that the parameters $R_p$ are generally greater but close to unity.

The coefficients $\Sigma_p$ are however not directly related to the $S_p$ coefficients. Similarly to those coefficients they also depend on the power spectrum index. The difference is that the $S_p$ parameters are averages of products of $N_p$ functions over the wave vectors, whereas the $\Sigma_p$ parameters are averages of the same products but over the *orthogonal* part of the wave vectors. This latter problem is similar to the one that have been solved to get the $S_p$. The only change is that the calculation has to be done in a 2D Fourier space instead of a 3D. To derive this series I follow the method developed by Bernardeau (1992, 1994b).

### 2.3.1. The $\Sigma_p$ parameters without smoothing

The first step of the calculation is to derive the generating function of the coefficient $\Sigma_p$ when one makes the approximation that $W_2(l) \approx 1$. It will turn out that it corresponds to the case $n = -2$. For 3D calculations the corresponding case has been considered by Bernardeau (1992). The only difference with the 3D case is that the mean value of $(k_1.k_2)^2/(k_1 k_2)^2$ is $1/2$ instead of $1/3$. The generating function of the vertices (i.e., the monopole part of equation,

$$-(1+\mathcal{G}_\delta)\tau^2 \frac{d^2}{d\tau^2}\mathcal{G}_\delta + \frac{3}{2}\left(\tau\frac{d}{d\tau}\mathcal{G}_\delta\right)^2$$
$$-\frac{3}{2}(1+\mathcal{G}_\delta)\tau\frac{d}{d\tau}\mathcal{G}_\delta + \frac{3}{2}\mathcal{G}_\delta(1+\mathcal{G}_\delta)^2 = 0. \quad (27)$$

with $\mathcal{G}_\delta(\tau) \sim -\tau$ when $\tau \to 0$,

instead of equation (27) of Bernardeau (1992). I did not find any simple solution for this differential equation, but it can be shown that it corresponds to the equation describing the "spherical" collapse in 2D, where $\mathcal{G}_\delta(\tau)$ is the density contrast and $-\tau$ is the linear density contrast. It can also be shown that

$$\mathcal{G}_\delta(\tau) - 1 \sim \tau^{-(\sqrt{13}-1)/2} \quad \text{when} \quad \tau \to \infty \quad (28)$$

and that the form

$$\mathcal{G}_\delta(\tau) = \left(1 + \frac{\tau}{\nu}\right)^{-\nu} - 1 \quad \text{with} \quad \nu = \frac{\sqrt{13}-1}{2} \quad (29)$$

provides a good fit for the solution of equation (27). More precisely one can rigorously calculate the expansion of $\mathcal{G}_\delta(\tau)$ near $\tau = 0$ and it reads

$$\mathcal{G}_\delta(\tau) = -\tau + \frac{12}{14}\tau^2 - \frac{29}{42}\tau^3 + \frac{79}{147}\tau^4 - \frac{2085}{5096}\tau^5 + \ldots \quad (30)$$

The generating function of the parameters $\Sigma_p$,

$$\varphi(y) = \sum_{p=1}^{\infty} \Sigma_p \frac{(-y)^p}{p!}, \quad \Sigma_1 = \Sigma_2 = 1, \quad (31)$$

is solution of the system,

$$\varphi(y) = y + y\mathcal{G}_\delta[\tau(y)] + \frac{1}{2}\tau(y)^2 \quad (32)$$
$$\tau(y) = -y\mathcal{G}'_\delta[\tau(y)].$$

The system (32) is a standard result of tree summation (Bernardeau & Schaeffer 1992, Bernardeau 1992). Using the expansion (30) for the function $\mathcal{G}_\delta(\tau)$ one can calculate the expansion of $\varphi(y)$ to get the first values of $\Sigma_p$. I found,

$$\begin{aligned}
\Sigma_3(n=-2) &= \frac{36}{7}; \\
\Sigma_4(n=-2) &= \frac{2540}{49}; \\
\Sigma_5(n=-2) &= 793; \\
\Sigma_6(n=-2) &= 16370; \\
&\ldots
\end{aligned} \quad (33)$$

These results do not take into account the filtering effects and are then independent of the shape of the power spectrum. They are actually exact for $n = -2$ since the integrales are then dominated by the $\mathbf{k} \to 0$ limit where $W_2(x\, k\, \theta_0) \approx 1$.

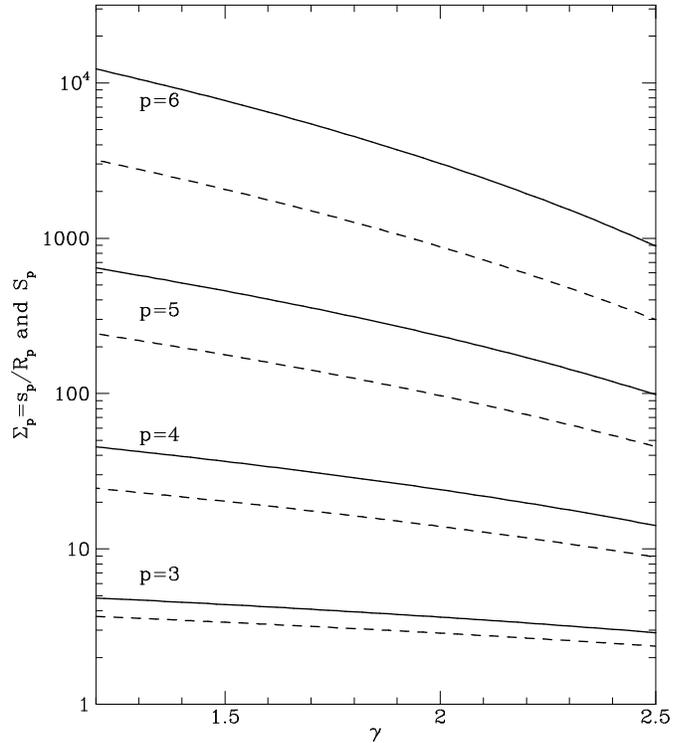

**Fig. 1.** The ratios $\Sigma_p$ as a function of the power law index, $\gamma$ (solid lines) for $p = 3, \ldots, 6$ and for an Einstein-de Sitter Universe. The ratios $S_p$ (Eq. [1]) for 3D statistics are represented by dashed lines.

### 2.3.2. The exact $\Sigma_p$ series

The derivation of the exact $\Sigma_p$ series has to take into account the geometrical dependences contained in the window function (i.e. in the last section I assumed actually that $W_2(|\mathbf{k}_1 + \ldots + \mathbf{k}_p|) = W_2(k_1)\ldots W_2(k_p)$). Calculations made in 3D prove that the effects of the filtering are significative. Comparison of $\Sigma_3$ in (33) and the result (22) proves also that it has to be taken into account.

For a top-hat window function the properties (20) and (21) show that the smoothing corrections lead to analytically simple corrections. It also demonstrates that we encounter the same situation than for the derivation of the smoothing corrections in 3D with a top-hat window function. Following the structure of the results obtained by Bernardeau (1994b) I expect that the smoothing corrections will be given by a Lagrangian-Eulerian space transformation. The generating function of the exact cumulants, $\varphi(y)$, is then given by a system similar to (32) but with a generating function of vertices $\mathcal{G}^S_\delta(\tau)$ given by

$$\mathcal{G}^S_\delta(\tau) = \mathcal{G}_\delta\left(\tau\left[1 + \mathcal{G}^S_\delta(\tau)\right]^{-(2+n)/4}\right). \quad (34)$$

then quite straightforward and can be done easily with a mathematical symbolic package.

I give here the expression of the first few parameters $\Sigma_p$,

$$\Sigma_3 = \frac{36}{7} - \frac{3(n+2)}{2};$$
$$\Sigma_4 = \frac{2540}{49} - 33(n+2) + \frac{21(n+2)^2}{4};$$
$$\Sigma_5 = 793 - 794(n+2) + 265(n+2)^2 \quad (35)$$
$$\qquad - 29.4(n+2)^3;$$
$$\Sigma_6 = 16370 - 22511(n+2) + 11594(n+2)^2$$
$$\qquad - 2650(n+2)^3 + 226.9(n+2)^4;$$
$$\ldots$$

In general there are no simple closed forms for the ratios $\Sigma_p$.

In Fig. 1, I present the dependence of the ratios $\Sigma_p$ as a function of the power law index. They are compared to the $S_p$ parameters (Eq. [1]) for the 3D statistics (dashed lines). It shows that they are significantly greater than their 3D counterparts.

### 2.4. The $\Omega$ dependence of the $s_p$ parameters

In the small angle limit the factorisation property (25) is always valid, whatever the value of $\Omega$. The parameters $s_p$ can then depend on $\Omega$ through the parameters $R_p$ or the parameters $\Sigma_p$. The $\Omega$ dependence of $R_p$ appears only when the time evolution of the correlation functions is taken into account (that is for very deep surveys), since the radius-time relationship is $\Omega$ dependent. But as in all the results that have been presented in this paper such a time dependence has been ignored, the $\Omega$ dependence of $R_p$ will not be discussed furthermore and be assumed negligible.

The $\Omega$ dependence is then entirely contained in the $\Sigma_p$ parameters. To solve this problem in general we have to use the dynamics of the "2D spherical collapse" for any value of $\Omega$, which is not known. It is possible however to get, without too much difficulty, the value of $\Sigma_3$ in the $\Omega \to 0$ limit. This is based on the result that in such a limit we have (see Bouchet et al. 1992),

$$\frac{\rho^{(2)}(\boldsymbol{x})}{\rho_0}(\Omega = 0) = \int \frac{\mathrm{d}^3\boldsymbol{k}_1}{(2\pi)^{3/2}} \frac{\mathrm{d}^3\boldsymbol{k}_2}{(2\pi)^{3/2}} \, \delta(\boldsymbol{k}_1) \, \delta(\boldsymbol{k}_2)$$
$$\times \, \exp[\mathrm{i}(\boldsymbol{k}_1 + \boldsymbol{k}_2)\boldsymbol{x}] \left[\frac{3}{4} + \frac{\boldsymbol{k}_1.\boldsymbol{k}_2}{k_2^2} + \frac{1}{4}\left(\frac{\boldsymbol{k}_1.\boldsymbol{k}_2}{k_2 \, k_1}\right)^2\right]. \quad (36)$$

The resulting value of $\Sigma_3$ is then

$$\Sigma_3(\Omega = 0) = \frac{21}{4} - \frac{3(n+2)}{2}. \quad (37)$$

It demonstrates that for $\Sigma_3$ the $\Omega$ dependence is extremely weak: between $\Omega = 1$ and $\Omega \to 0$ the variation of $\Sigma_3$ is only of the order of 3%. More precisely following Bouchet et al. (1992) one can use an analytical fit to describe the $\Omega$ dependence of $\Sigma_3$ valid for $0.05 \lesssim \Omega \lesssim 3.$,

$$\Sigma_3(\Omega) = \frac{36}{7} + \frac{9}{14}\left(\Omega^{-2/63} - 1\right) - \frac{3(n+2)}{2}. \quad (38)$$

A similar weak $\Omega$ dependence is expected for the higher order parameters $\Sigma_p$ (as it is the case for the $S_p$ coefficients).

## 3. Validity of the small angle approximation

The expressions (22-23) of the angular skewness and the higher order coefficients (26, 35) rely on the approximation (15) which corresponds to the small angle limit. The validity domain of this approximation is however questionable. This investigation is all the more crucial that in principle the quasilinear regimes applies at large physical scales which would correspond to rather large smoothing angles.

To test the validity of this approximation I computed the $s_3$ parameter from direct Monte-Carlo numerical integration for different smoothing angles and sample characteristics.

### 3.1. Model for galaxy clusters

To start with I used a very simple model given by,

$$\xi_2^{\mathrm{model}}(r) = \left(\frac{r}{r_0}\right)^{-\gamma}, \quad (39)$$

with $\gamma = 1.5$, 1.7 or 1.9, and $r_0 = 20h^{-1}\mathrm{Mpc}$. The selection function is simply assumed to be a Heavyside function,

$$F^{\mathrm{model}(1)}(r) = 1 \text{ for } r < D \quad (40a)$$

and

$$F^{\mathrm{model}(1)}(r) = 0 \text{ for } r > D \quad (40b)$$

with $D \approx 600h^{-1}\mathrm{Mpc}$. This could actually be considered as a good parametrization of the properties of galaxy cluster samples (Tóth et al. 1989). The other advantage of such a choice is that it allows to test the Monte-Carlo integration against analytical results in two cases: in the small angle limit, and for $\theta_0 = 180^\circ$, the latter corresponding to the filtering with a spherical top-hat window function. (it has obviously no practical interest for angular surveys!)

The results for the mean 2-point angular function are presented in Fig. 2. The solid lines show the results of Monte-Carlo integrations for different angles and different values of $\gamma$. The dashed lines correspond to the small angle Limber approximations,

$$\overline{w}_2(\theta_0) = \left(\frac{r_0}{D}\right)^\gamma \theta_0^{1-\gamma} \frac{9\,\Gamma(\frac{3-\gamma}{2})\,\Gamma(2-\frac{\gamma}{2})\,\Gamma(\frac{-1+\gamma}{2})}{2^{\gamma-4}(6-\gamma)(3-\gamma)\Gamma(\frac{7-\gamma}{2})\Gamma(\frac{\gamma}{2})}, \quad (41)$$

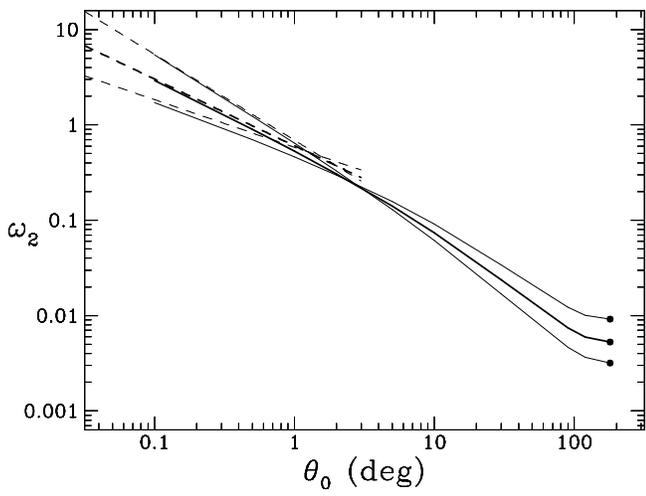

**Fig. 2.** The mean 2-point angular correlation function as a function of the smoothing angle for the model (1) (39, 40). The thick solid line is for $\gamma = 1.7$ and the thin solid lines for $\gamma = 1.5$ and $\gamma = 1.9$. The dashed lines correspond to the small angle limit (41) and the circle to the $\theta_0 = 180^\circ$ case (eq. [42]).

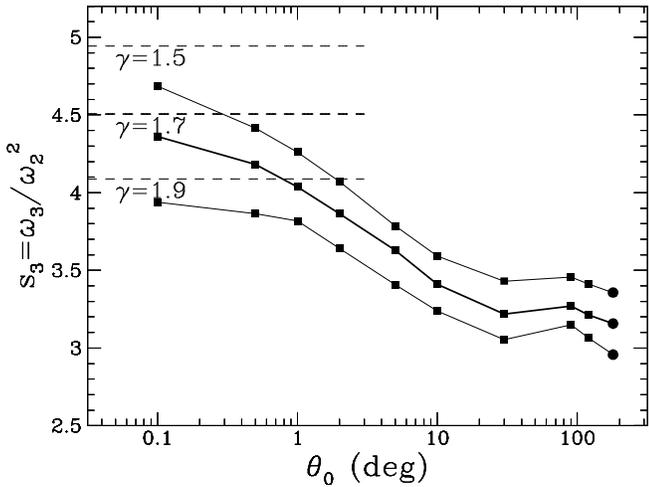

**Fig. 3.** The value of $s_3$ as a function of the smoothing angle for the model (1) (39,40). The notations are the same than in Fig. 2 where the dashed lines correspond here to the small angle limit value, (44) and the circles to the expression, $S_3 = 34/7 - \gamma$.

and the circle to the exact integration corresponding to the case of the spherical top-hat window function

$$\overline{\omega}_2(180^\circ) = \left(\frac{r_0}{D}\right)^\gamma \frac{9 \, 2^{3-\gamma}}{(6-\gamma)(\gamma-4)(\gamma-3)}. \tag{42}$$

The mean three-point angular correlation function has been obtained using the expression of the three-point correlation function in real space given by the second order perturbation theory (Fry 1984),

$$\xi_3(\boldsymbol{r}_1, \boldsymbol{r}_2, \boldsymbol{r}_3) = \left[\frac{10}{7} - \frac{\gamma}{3-\gamma} \cos(\theta_1) \left(\frac{r_{21}}{r_{31}} + \frac{r_{31}}{r_{21}}\right)\right.$$
$$\left. + \frac{4}{7} \frac{3 - 2\gamma + \gamma^2 \cos^2\theta_1}{(3-\gamma)^2}\right] \xi_2(\boldsymbol{r}_1, \boldsymbol{r}_2) \, \xi_2(\boldsymbol{r}_1, \boldsymbol{r}_3) \tag{43}$$
$$+ \text{cyc.}(1,2,3)$$

where $\theta_1$ is the angle between the vectors $\boldsymbol{r}_2 - \boldsymbol{r}_1$ and $\boldsymbol{r}_3 - \boldsymbol{r}_1$ and $r_{21}$ and $r_{31}$ are the lengths of these vectors. For the parameters of this model the equations (22, 23) give, for an Einstein-de Sitter Universe[1],

$$s_3^{model(1)}(\theta_0 \to 0) = \frac{(6-\gamma)^2}{3(9-2\gamma)} \left(\frac{93}{14} - \frac{3}{2}\gamma\right). \tag{44}$$

The resulting values for $s_3$ ($\equiv \overline{\omega}_3/\overline{\omega}_2^2$) are presented in Fig. 3 as a function of $\theta_0$ for the different values of $\gamma$. The notations are the same than in Fig. 2. It has to be noticed that the small angle limit is reached only for very small angle, e.g. significantly smaller than $1^\circ$. The transition between the small angle result and the $\theta_0 = 180^\circ$ case happens for $\theta \approx 1^\circ$.

It implies that the analytical results of the previous part cannot be applied exactly to the observations in the $1^\circ \sim 5^\circ$ range for such a model. However this result suggests that the exact values are intermediate between the small angle limits and the value of the $S_p$ parameters (Fig. 1). It thus gives a plausible range where the exact values of the $s_p$ parameters are expected to be.

### 3.2. Model for galaxies

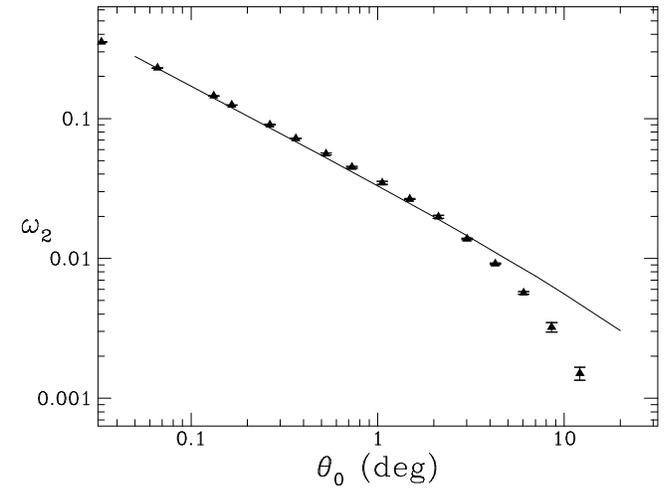

**Fig. 4.** The mean 2-point angular correlation function as a function of the smoothing angle for the model (2), eq. (45), of the selection function and for $\gamma = 1.7$. The triangles correspond to the observations in the APM galaxy survey (Gaztañaga 1994)

---

[1] In the limit $\Omega \to 0$ the factor $(93/14 - 3\gamma/2)$ becomes $(27/4 - 3\gamma/2)$.

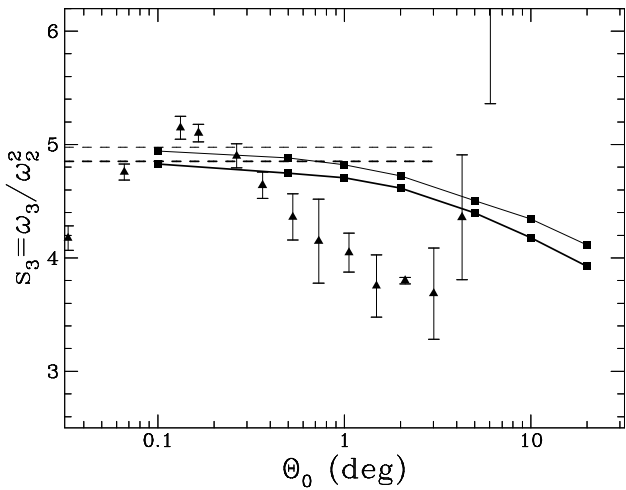

**Fig. 5.** The value of $s_3$ as a function of the smoothing angle for the model (2). The dashed lines correspond to the small angle limits; the thin lines to the $\Omega \to 0$ case; the thick lines to the $\Omega = 1$ case. The triangles correspond to the observations in the APM galaxy survey (Gaztañaga 1994)

The second model I considered has been built to mimic the properties of the galaxy angular surveys. The selection function I used is then of the form

$$F^{\mathrm{model}(2)}(r) \propto r^{-0.5} \exp\left[-\left(\frac{r}{D}\right)^2\right]. \qquad (45)$$

For such a model the small scale limit of the coefficient $s_3$ is, for an Einstein-de Sitter Universe,

$$\begin{aligned}s_3^{\mathrm{model}(2)}(\theta_0 \to 0) =& \frac{32}{27} \frac{\Gamma(5/4)}{3^{3/4}} \left(\frac{3}{2}\right)^\gamma \frac{\Gamma\left(\frac{15}{4} - \gamma\right)}{\Gamma^2\left(\frac{5-\gamma}{2}\right)} \\ & \times \left(\frac{93}{14} - \frac{3}{2}\gamma\right).\end{aligned} \qquad (46)$$

For the 2-point correlation function I assumed a power law behaviour with $\gamma = 1.7$ and $r_0 = 5h^{-1}$Mpc. It reproduces reasonably well the observed angular mean two-point correlation function (Fig. 4.) in the APM galaxy survey for $\theta_0 \lesssim 2^\circ$. The resulting $s_3$ parameters are given in Fig. 5. It can be seen that the small angle approximation is slightly better than for the previous case. It is not too surprising since the boundaries are in this case smoother, so that the small angle approximation (which basically neglects the effects of boundaries for 2 points out of 3) is expected to be more accurate. In such a case the values found at the $1^\circ$ scale are in reasonable agreement with the small scale limit, although slightly lower.

### 4. Comparison with observations and conclusions

The measures of such $s_p$ coefficients have been made in the APM survey by Gaztañaga (1994) up to $p = 9$ in the $\theta_0 = 0.05 \sim 5$ range. For a depth of about $350h^{-1}$Mpc the expected validity domain of the perturbation theory is for $\theta_0 \gtrsim 1^\circ$ (which corresponds to a real scale greater $5h^{-1}$Mpc in which perturbation theory has been proved valid, Bernardeau 1994b, Juszkiewicz et al. 1994, Baugh et al. 1994).

When they are compared to the observations the perturbation theory predictions for $s_3$ in Fig. 5 are found to be greater (4.6 instead of $3.8 \pm 0.2$ for $\theta_0 \approx 2^\circ$) in the $1 - 5^\circ$ range. This excess is only about 20% but leads to a significative disagreement between perturbation theory and the observations, according to the errorbars claimed by Gaztañaga, and contrary to the conclusions of the analysis made by Frieman & Gaztañaga (1994). However, it is important to have in mind that the perturbation theory predictions obtained in the previous section neglected effects such as the redshift evolution of the correlation functions, the departure of the two-point correlation function from a power law bahaviour, errorbars for the determination of the 2-point correlation function. These effects may change the results. Thus more precise studies have to be undertaken before any definitive conclusion can be given.

**Table 1.** The $s_p$ parameters: theory and observation

| $p$ | $R_p$ | $s_p(\theta \to 0)$ | obs: $s_p(\theta \gtrsim 1^\circ)$ |
|---|---|---|---|
| 3 | 1.19 | 4.87 | $3.81 \pm 0.07$ |
| 4 | 1.52 | 47.6 | $32.5 \pm 4.2$ |
| 5 | 2.00 | 713.5 | $384 \pm 62$ |
| 6 | 2.71 | 14730 | $3260 \pm 1340$ |

In table 1 I give the expected values of the $s_p$ ($p = 3, 6$) coefficients in the small angle limit using the parameters $R_p$ of Gaztañaga (1994) for $\gamma = 1.7$. The actual values predicted by the perturbation theory are expected to be close to these numbers when $\theta_0 \sim 1^\circ$. The observations are seen to be systematically smaller than the predictions, similarly to the $p = 3$ case. These predictions, however, rely entirely on the small angle approximation.

If we admit that the discrepancy between perturbation theory results and observations is real, what would it mean? One could argue that perturbation theory does not apply, that nonlinear corrections affect the result. I think that this is unlikely to be the case since the main contribution to the averaged angular correlation functions comes from galaxies being at a distance of about $\theta_0 D$ which for the $1^\circ$ scale is large enough for the perturbation theory to apply. Note also that the nonlinear corrections are rather expected to increase the values of the $s_p$ parameters and thus amplify the discrepancy. A more serious concern would be the hypothesis that the galaxies are not biased. In this case it does not mean only that the so called $b$ parameter should be 1, more constrainly it means that the density contrast observed in Galaxy counts re-produces the *nonlinear* aspects of the matter density contrast (see for instance the discussion by Fry & Gaztañaga 1993). Even if most of the mass of the universe is in galaxies (possibly in haloes), the simple fact that there might be a segregation in luminosity for the correlation function (Hamilton 1988, Park et al. 1994, Loveday et al. 1994 but see also Alimi, Valls-Gabaud & Blanchard 1988 for a different conclusion) is likely to affect the whole statistics. The basic reason is that the various contributions are not properly weighted (i.e. by mass) when the correlation functions are measured. Bernardeau & Schaeffer (1992) explored the consequences of such effects on the two- and three-point correlation functions in the frame of a gravitationally induced luminosity segregation. They found significative changes for the value of $S_3$. In particular, for the limit of rare objects (either very massive clusters or very bright galaxies) the value of $S_3$ is expected to be close to 3 for any cosmological models. This limit is not likely to be relevant for galaxy surveys but may provide a good model for the observed galaxy cluster correlation functions.

In any case, even if the agreement between the observations and the perturbation theory results is not perfect, it supports the idea that the present large–scale structures have been built out by the groth of initial Gaussian fluctuations in a gravitational instability scheme. The $s_p$ coefficients provide then a unique and powerful way to test the large–scale structure formation models for gravitational instability scenarios.

*Acknowledgements.* The author thanks D. Pogosyan, C. Pichon and A. Blanchard for useful discussions.